\documentclass[pdflatex,sn-mathphys-num]{sn-jnl}

\usepackage{graphicx}%
\usepackage{multirow}%
\usepackage{amsmath,amssymb,amsfonts}%
\usepackage{amsthm}%
\usepackage{mathrsfs}%
\usepackage[title]{appendix}%
\usepackage{xcolor}%
\usepackage{textcomp}%
\usepackage{manyfoot}%
\usepackage{booktabs}%
\usepackage{algorithm}%
\usepackage{algorithmicx}%
\usepackage{algpseudocode}%
\usepackage{listings}%

\usepackage{siunitx}
\usepackage[utf8]{inputenc}

\DeclareSIUnit{\ionsperGHz}{\text{ions}/\si{\giga\hertz}}
\DeclareSIUnit{\ionspercubecm}{\text{ions}/\si{\cubic\cm}}
\DeclareSIUnit{\ionsperkHz}{\text{ions}/\si{\kilo\hertz}}
\DeclareSIUnit{\ions}{ions}
\DeclareSIUnit{\angstrom}{\text{\AA}}

\theoremstyle{thmstyleone}%
\theoremstyle{thmstyletwo}%

\theoremstyle{thmstylethree}%

\begin{document}

\title[Optically detected nuclear magnetic resonance of
coherent spins in a molecular complex]{Optically detected nuclear magnetic resonance of
coherent spins in a molecular complex}


\author*[1,2]{\fnm{Evgenij} \sur{Vasilenko}}\email{evgenij.vasilenko@kit.edu}
\equalcont{These authors contributed equally to this work.}

\author[2]{\fnm{Vishnu} \sur{Unni Chorakkunnath}}\email{vishnu.chorakkunnath@kit.edu}
\equalcont{These authors contributed equally to this work.}

\author[2]{\fnm{Jeremias} \sur{Resch}}

\author[2]{\fnm{Nicholas} \sur{Jobbitt}}

\author[3]{\fnm{Diana} \sur{Serrano}}

\author[3]{\fnm{Philippe} \sur{Goldner}}

\author[1]{\fnm{Senthil Kumar} \sur{Kuppusamy}}

\author[1,4,5]{\fnm{Mario} \sur{Ruben}}

\author*[1,2]{\fnm{David} \sur{Hunger}}\email{david.hunger@kit.edu}

\affil*[1]{\orgdiv{Institute for Quantum Materials and technologies (IQMT)}, \orgname{Karlsruhe Institute of Technology}, \orgaddress{\city{Karlsruhe}, \postcode{76131},\country{Germany}}}

\affil*[2]{\orgdiv{Physics Institute (PHI)}, \orgname{Karlsruhe Institute of Technology}, \orgaddress{\city{Karlsruhe}, \postcode{76131},\country{Germany}}}

\affil[3]{\orgdiv{Chimie ParisTech}, \orgname{PSL University, CNRS, Institut de Recherche de Chimie Paris}, \orgaddress{\city{Paris}, \postcode{75231}, \country{France}}}

\affil[4]{\orgdiv{Institute of Nanotechnology (INT)}, \orgname{Karlsruhe Institute of Technology}, \orgaddress{\city{Karlsruhe}, \postcode{76131}, \country{Germany}}}

\affil[5]{\orgdiv{Centre Européen de Sciences Quantiques (CESQ)}, \orgname{Institut de Science et d'Ingénierie Supramoléculaires (ISIS)}, \orgaddress{\city{Strasbourg}, \postcode{67083}, \country{France}}}


\abstract{Nuclear magnetic resonance (NMR) is a powerful tool for applications ranging from chemical analysis to quantum information processing. Achieving optical initialization and detection of molecular nuclear spins promises new opportunities — including improved NMR signals at low magnetic field, sensitivity down to the single-molecule level, and full access to atomically precise molecular architectures for quantum technologies. In this study, we report optical readout of coherently controlled nuclear spins in a europium-based molecular crystal. By harnessing ultra-narrow optical transitions, we achieve optical initialization and detection of nuclear spin states. Through radio-frequency driving, we address two nuclear quadrupole resonances, characterized by narrow inhomogeneous linewidths and a distinct correlation with the optical transition frequency. We implement Rabi oscillations, spin echo and dynamical decoupling techniques, achieving nuclear spin quantum coherence with a lifetime of up to \SI{2}{\milli\second}. These results highlight the capabilities of optically detected NMR (ODNMR) and underscore the potential of molecular nuclear spins for quantum information processing.}


\maketitle

\noindent
\section{Introduction}\label{sec:intro}
NMR is a well-established and highly developed field that plays a vital role across a wide range of applications - from pharmaceutical quality control to materials research. Due to their weak interaction with the environment, nuclear spins are at the same time a valuable resource for quantum technology \cite{awschalom_quantum_2018,marcks_nuclear_2025} that may allow for dense qubit registers which can be operated at comparably high temperature. Optical addressing of nuclear spins serves as an important tool to leverage them as qubits for quantum memories \cite{afzelius_multimode_2009,ma_one-hour_2021,bradley_robust_2022} and quantum processors \cite{waldherr_quantum_2014,abobeih_fault_2022}. 
Nuclear spin addressing is usually achieved only indirectly \cite{suter_optical_2020} through optical transitions linked to electron spins. Such addressing is successfully used in color centers in diamond \cite{childress_coherent_2006,waldherr_quantum_2014,abobeih_fault_2022,beukers_control_2025}, silicon carbide \cite{hesselmeier_high-fidelity_2024,zeledon_minute-long_2025}, and semiconductor quantum dots \cite{gammon_nuclear_1997,appel_many-body_2025}. 
Optically addressable molecular spin qubits \cite{kuppusamy_spin-bearing_2024,atzori_optical_2025} offer a novel platform in this context, where optical and spin properties can be tailored, photonic integration is facilitated \cite{toninelli_single_2021}, and supra-molecular assemblies may enable atomically precise qubit registers \cite{ruben_grid-type_2004}. Early pioneering work has introduced optically detected magnetic resonance in molecular spins down to the single molecule level \cite{kohler_magnetic_1993,wrachtrup_optical_1993}, and recent results based on transition metal complexes \cite{bayliss_optically_2020,bayliss_enhancing_2022}, organic radicals \cite{gorgon_reversible_2023,chowdhury_optical_2024}, and lanthanide complexes \cite{serrano_ultra_2022,kuppusamy_observation_2023,weiss_high-resolution_2025} have spurred new interest. 

Indirect optical addressing of nuclear spins relies on magnetic coupling between electron and nuclear spins. This coupling is typically weak and thus constrains the bandwidth for spin initialization, manipulation, and readout. Furthermore, the presence of the electron spin typically introduces noise and limits the coherence of coupled nuclear spins \cite{maurer_room-temperature_2012}. The strong magnetic moment of electron spins also limits their useful density to avoid undesired couplings \cite{chiossi_optical_2024}.

Trivalent non-Kramers rare-earth ions such as Eu$^{3+}$ and Pr$^{3+}$ represent a notable exception by offering directly addressable nuclear spins. They possess no net electronic spin and offer narrow optical transitions that enable direct resolution, initialization, manipulation, and readout of nuclear spin states \cite{goldner_chapter_2015}. For example, in europium-doped yttrium orthosilicate (Eu$^{3+}$:$\mathrm{Y}_2\mathrm{Si}\mathrm{O}_5$) crystals, optically detected nuclear spin quantum coherence with a lifetime exceeding ten hours has been demonstrated \cite{wang_nuclear_2025}, and quantum storage of photons in nuclear spin states for up to one hour has been achieved \cite{ma_one-hour_2021}. 
Recently it has been shown that europium-based molecular complexes show outstanding optical coherence properties and long nuclear spin lifetimes, which allow for direct optical nuclear spin access and optical spin initialization \cite{serrano_ultra_2022,kuppusamy_observation_2023}. However, combining optical addressing and coherent nuclear spin control has remained elusive for molecular complexes so far.

\section{Results and discussion}\label{sec:results}

\begin{figure}[h]
	\centering
	\includegraphics[width=\textwidth]{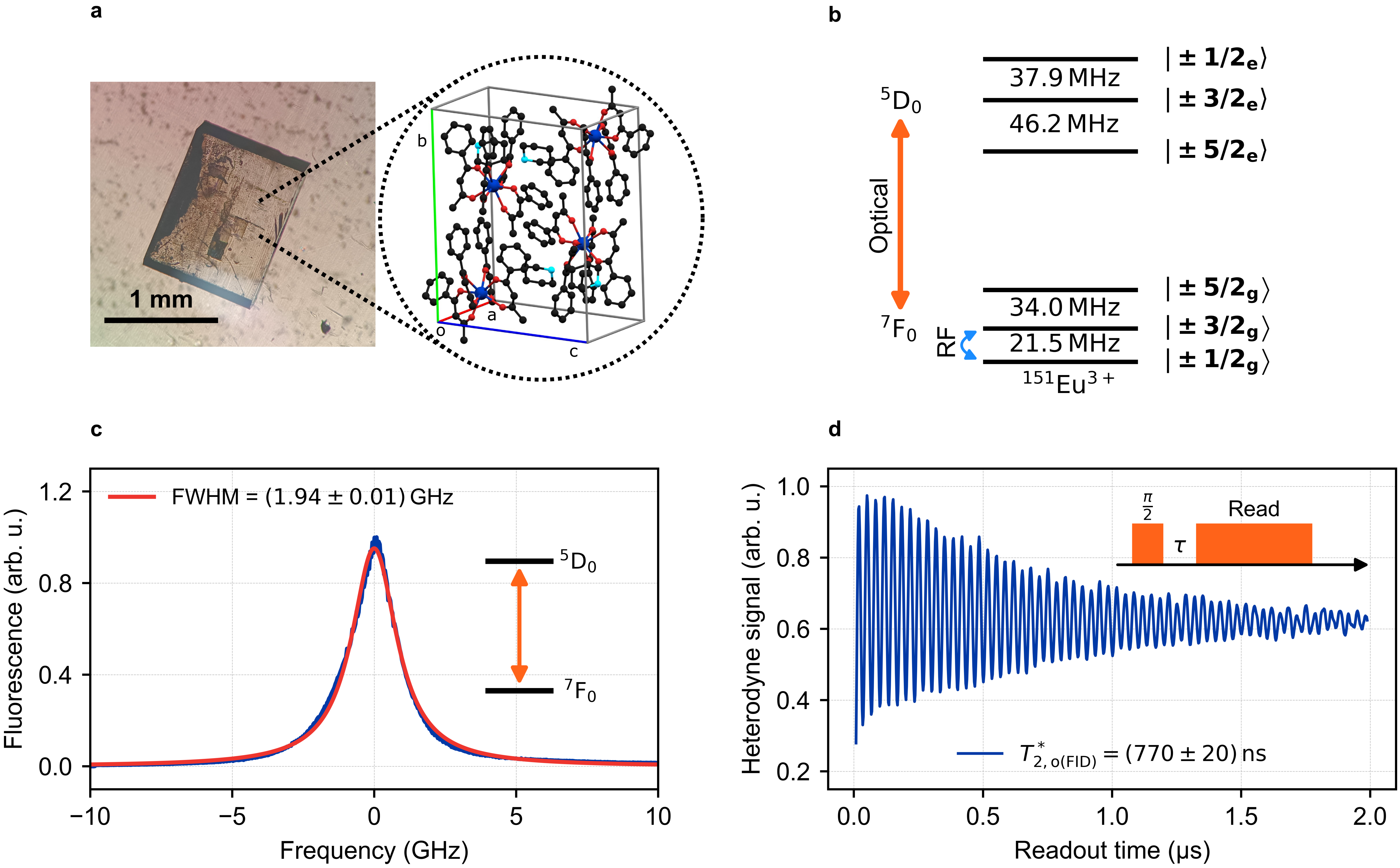} 
	\caption{\textbf{Molecular crystal and optical properties. (a)} Molecular structure of the complex and crystal unit cell obtained from single-crystal X-ray diffraction. In blue: europium, in red: oxygen, in black: carbon, in cyan: nitrogen; hydrogen is omitted for clarity. Single crystals with \SI{\sim 1}{\milli\meter} size and clear facets were grown. \textbf{(b)} The ground and excited-state hyperfine levels of the $\mathrm{}^7\mathrm{F}_0\rightarrow\mathrm{}^5\mathrm{D}_0$ transition of $^{151}\mathrm{Eu}^{3+}$. \textbf{(c)} Photoluminescence excitation measurement of the $\mathrm{}^7\mathrm{F}_0\rightarrow\mathrm{}^5\mathrm{D}_0$ transition showing a narrow inhomogeneous linewidth of \SI{1.94 \pm 0.01}{\giga\hertz}. \textbf{(d)} Optical free-induction decay of the optical transition dipoles of the ions. Coherent oscillations can be observed as a beating signal via heterodyne detection with a frequency-detuned optical readout pulse. A fit to the signal yields a pure dephasing time $T^*_{2,\text{o(FID)}}=$ \SI{770 \pm 20}{\nano\second}.} 
	\label{fig:figure1} 
\end{figure}

In this work, we demonstrate optical initialization and readout of coherently controlled nuclear spins in a europium-based molecular complex, marking a significant advance towards establishing this system as a viable platform for quantum technologies. We study stoichiometric molecular crystals composed of a mononuclear $\mathrm{Eu}^{3+}$ complex $[\mathrm{Eu(BA)_4}\mathrm{(pip)}]$, where $\mathrm{BA}$ and $\mathrm{pip}$ refer to benzoylacetonate and piperidinium, respectively. The complex is constituted of an eight-coordinated anionic fragment $[\mathrm{Eu(BA)_4}]^-$, and the charge balancing piperidinium cation. The complex crystallizes in a monoclinic lattice with four molecules per cell as shown in Fig.~\ref{fig:figure1}a. A first detailed optical characterization of the europium complex in a microcrystalline powder \cite{serrano_ultra_2022} has evidenced narrow optical homogeneous linewidths connected to the $^{7}\mathrm{F}_{0}\rightarrow\mathrm{}^{5}\mathrm{D}_{0}$ transition, as well as optical spin polarization and long-lived nuclear spin states, which have allowed to infer the energy level structure, see Fig.~\ref{fig:figure1}b. To optimize the homogeneity of the material, we have grown millimeter-sized molecular crystals via slow solvent evaporation, see Fig.~\ref{fig:figure1}a and Supplementary Materials. For optical readout, we incorporate a crystal into a fiber-based ferrule setup (see Supplementary Materials). The setup is directly immersed in liquid helium, providing effective thermalization and a stable temperature of \SI{4.2}{\kelvin}. In addition, a superconducting coil is installed to address the hyperfine transitions by applying radio-frequency (RF) fields. 

As a first step, we characterize the optical properties to assess the impact of crystal quality. 
We scan a tunable dye laser across the $^{7}\mathrm{F}_{0}\rightarrow\mathrm{}^{5}\mathrm{D}_{0}$ transition and observe an inhomogeneous linewidth of $\Gamma_\text{inh}=\SI{1.94 \pm 0.01}{\giga\hertz}$, see Fig.~\ref{fig:figure1}c. This is a factor of 3 narrower compared to the value (\SI{6.6}{\giga\hertz}) reported for a micro-crystalline powder in a previous study \cite{serrano_ultra_2022}, evidencing a reduced amount of structural defects in the crystal. 
In order to access the homogeneous linewidth $\Gamma_\text{h}$, we performed spectral hole burning (SHB) measurements at low laser powers of \SI{20}{\micro\watt} to avoid power broadening. The narrowest observed hole width is \SI{620 \pm 60}{\kilo\hertz} (see Supplementary Materials), corresponding to a homogeneous linewidth $\Gamma_\text{h}=\SI{310}{\kilo\hertz}$ and a coherence time $T^*_2=1/(\pi\Gamma_\text{h})= \SI{1.03 \pm 0.10}{\micro\second}$. 

While SHB probes the long-term homogeneous linewidth, a measurement of the optical free-induction decay (FID) gives access to the instantaneous pure dephasing time $T^*_{2,\text{o}}$ \cite{devoe_subnanosecond_1979}. The pulse sequence consists of an optical $\pi/2$ pulse that creates an optical coherence and a frequency-shifted readout pulse that interferes with the radiated field for heterodyne detection. The decaying beating signal yields a lower limit for the dephasing time of $T^*_{2,\text{o(FID)}}= \SI{0,77 \pm 0,020}{\micro\second}$, see Fig.~\ref{fig:figure1}d. This is slightly shorter compared to the SHB measurement and shows that the elevated power required for the coherent $\pi/2$ pulse in the FID measurement and thus power broadening is more dominant than spectral diffusion \cite{devoe_subnanosecond_1979}. To probe the optical coherence time $T_{2,\text{o}}$, two-pulse photon echo experiments with heterodyne detection were performed that yield $T_{2,\text{o}}= \SI{2,13 \pm 0.03}{\micro\second}$ (see Supplementary Materials). This value is also improved compared to the reported coherence time (\SI{1.49}{\micro\second}) at this temperature level \cite{serrano_ultra_2022}. Overall, this demonstrates that high-quality single crystals lead to improved inhomogeneous and homogeneous optical linewidths compared to micro-crystalline samples. 

\begin{figure} 
	\centering
	\includegraphics[width=\textwidth]{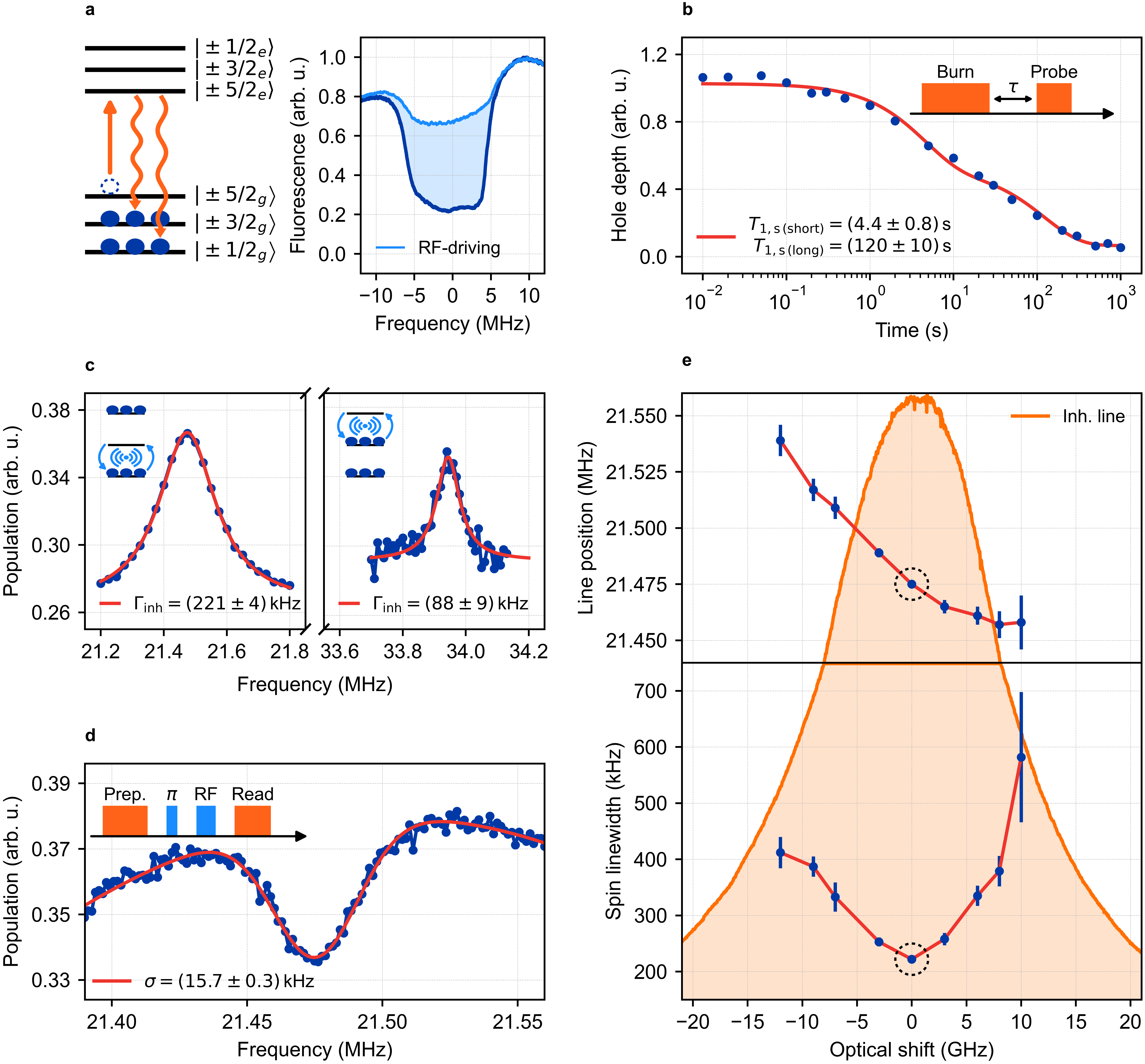} 
	\caption{\textbf{Optically detected nuclear magnetic resonance and spin state lifetime. (a)} Schematic illustration of depopulation of a hyperfine level by optical pumping for spin state preparation (left). A spectral pit of \SI{10}{\mega\hertz} width is prepared (right, dark blue), and resonant RF driving repopulates the level and fills the pit (light blue). \textbf{(b)} The time evolution of the pit depth is fitted with a double exponential decay, resulting in two spin relaxation times $T_{1,\text{s(short)}}=\SI{4.4}{\second}$ and $T_{1,\text{s(long)}}=\SI{122}{\second}$. \textbf{(c)} Pointwise measurement of an ODNMR spectrum for the two ground state transitions at \SI{21.475}{\mega\hertz} (left) and \SI{33.944}{\mega\hertz} (right). The spin transition has an inhomogeneous linewidth of \SI{221 \pm 4}{\kilo\hertz} and \SI{88 \pm 9}{\kilo\hertz}, respectively. \textbf{(d)} Spin hole burning spectrum to probe a homogeneous class of the spin transition. An RF $\pi$ pulse is applied to burn a spin hole before the ODNMR measurement is performed. The observed hole exhibits a linewidth of \SI{15.70 \pm 0.3}{\kilo\hertz}. \textbf{(e)} Measurement of the center frequency (top) and the spin inhomogeneous linewidth (bottom) of the \SI{21,5}{\mega\hertz} as a function of the optical frequency. The optical inhomogeneous line is shown in orange for comparison.}
	\label{fig:figure3}
\end{figure}

We now turn to optical spin initialization and detection of spin resonances. Therefore, we perform optical pumping to prepare a spectral pit, which results in the depletion of population of nuclear spin levels. While preparation of a sub-ensemble within a single hyperfine state can be achieved with a series of optical pulses \cite{nilsson_hole-burning_2004}, we find that sufficient signal contrast is achieved by burning a single spectral pit of \SI{10}{\mega\hertz} width with a laser chirp, which depopulates one hyperfine level for a certain ion class, see Fig.~\ref{fig:figure3}a. This is achieved faster than full class preparation, and is used in all subsequent experiments for initial spin state preparation. 
We measure the spin lifetime $T_{1,\text{s}}$ by probing the depth of the pit as a function of the waiting time. Fig.~\ref{fig:figure3}b shows the decay of the pit depth over time, yielding two characteristic time constants $T_{1,\text{s(short)}}=\SI{4,4 \pm 0,8}{\second}$ and $T_{1,\text{s(long)}}=\SI{120 \pm 10}{\second}$. Notably, $T_{1,\text{s(short)}}$ is one order of magnitude longer than reported previously \cite{serrano_ultra_2022}, despite higher temperature (4.2 vs. \SI{1.5}{\kelvin}). 

We focus on the ground-state nuclear quadrupole transitions of the isotope $^{151}\mathrm{Eu}^{3+}$ within a natural abundance sample, where the transition frequencies were previously estimated from SHB \cite{serrano_ultra_2022}, see Fig.~\ref{fig:figure1}b. We perform ODNMR for a precise determination of the transition frequencies and to probe the spin inhomogeneity. Therefore, after spin polarization, a weak optical probe pulse is applied, where we sweep the laser frequency by \SI{15}{\mega\hertz} over the spectral pit in order to read out the resulting fluorescence signal as a reference. Afterwards, a \SI{1}{\milli\second} RF pulse at a constant frequency with a power of \SI{\sim 92}{\watt} is applied with the coil. The resulting change of the spin population is read out with a final optical pulse to measure the fluorescence in the middle of the pit. We average the signal over five repetitions, and subsequently change the RF frequency before repeating the sequence to probe the spin transition point by point. Fig.~\ref{fig:figure3}c shows both ground-state resonances with center frequencies of \SI{33.944 \pm 0.002}{\mega\hertz} ($|3/2\rangle \leftrightarrow |5/2\rangle$) and \SI{21.475 \pm 0.001}{\mega\hertz} ($|1/2\rangle \leftrightarrow |3/2\rangle$), respectively.

The spin inhomogeneous line at \SI{34}{\mega\hertz} has a Lorentzian shape with a full width at half maximum (FWHM) of \SI{88}{\kilo\hertz}, similar to high-quality europium-doped solid state crystals \cite{goldner_chapter_2015,Arcangeli2014,serrano_all-optical_2018}, and almost a factor of three narrower than the \SI{21.5}{\mega\hertz} transition. The latter transition shows a much larger signal contrast for the same pulse parameters, indicating a larger transition strength. We observe a power-dependent broadening of the \SI{21,5}{\mega\hertz} transition and find a linewidth of \SI{154}{\kilo\hertz} at the lowest power of \SI{0,5}{\watt}.
We can estimate the homogeneous spin linewidth by performing spin hole burning. Therefore, we use the sequence for ODNMR and apply an additional RF $\pi$ pulse to remove a resonant class of spins from the probed ensemble. This produces a narrow hole in the spin inhomogeneous line (see Fig.~\ref{fig:figure3}d), and a Gaussian fit yields a hole width of \SI{15,70 \pm 0,32}{\kilo\hertz}, which reduces to \SI{11,64 \pm 0,60}{\kilo\hertz} when minimizing the RF burning power, corresponding to a lower bound of the spin dephasing time $T^*_{2,\text{s}}= \SI{19.3}{\micro\second}$.

The crystal used for the spin characterization has an optical inhomogeneous linewidth of \SI{23}{\giga\hertz}, indicating larger strain in the crystal, possibly due to mechanical forces during insertion into the ferrule setup. It is thus interesting to investigate the dependence of the spin transition properties across the optical inhomogeneous line. We therefore measure the line position and the inhomogeneous linewidth of the \SI{21,5}{\mega\hertz} transition, see Fig.~\ref{fig:figure3}e. We observe an approximately linear dependence of the spin transition frequency with the optical probing frequency with a gradient of \SI{-4}{\kilo\hertz\per\giga\hertz}. This shows an opposing sign and a smaller magnitude compared to the value \SI{10}{\kilo\hertz\per\giga\hertz} reported for solid state crystals \cite{yamaguchi_perturbed_1999,yamaguchi_three-parameter_2000}.
We furthermore observe a significant increase of the inhomogeneous broadening towards the wings of the optical line. This indicates that high crystalline quality also reflects in narrow spin inhomogeneous lines, and that strain affects optical and spin transitions in a correlated manner that is specific for the respective ligand field. These measurements exemplify the potential of ODNMR for studying materials properties, and for applications such as strain or pressure sensing \cite{singh_high_2024}.

\begin{figure} 
	\centering
	\includegraphics[width=\textwidth]{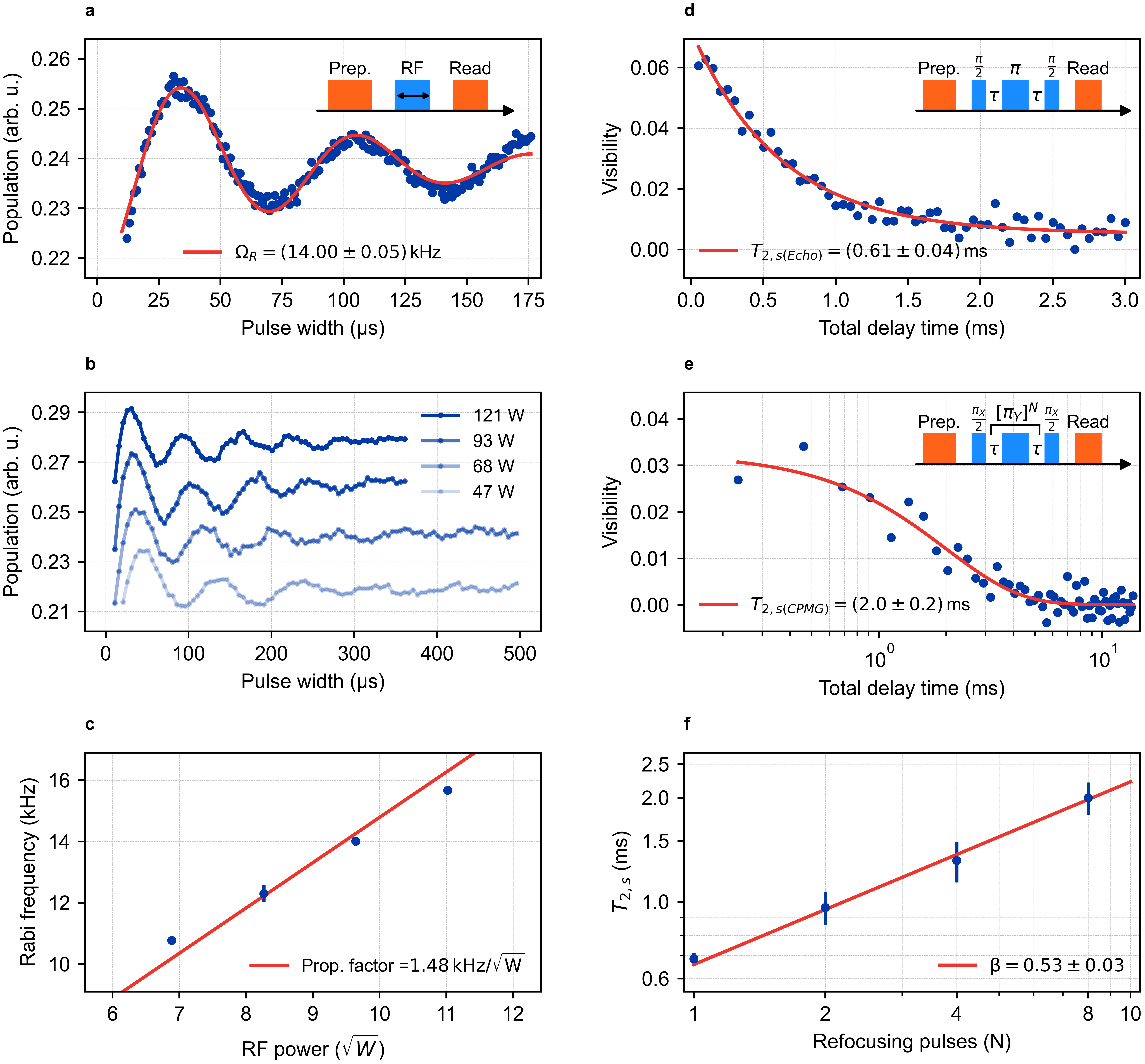} 
	\caption{\textbf{Coherent nuclear spin manipulation and spin coherence time. (a)} Rabi oscillations observed by varying the RF pulse duration, revealing a Rabi frequency of \SI{14}{\kilo\hertz}. The oscillations involve a damping component resulting from inhomogeneous broadening. \textbf{(b)} The Rabi frequency increases with increasing RF power. \textbf{(c)} The power dependence of the Rabi frequency shows the expected scaling with a proportionality factor of \SI{1.48}{\kilo\hertz\per\sqrt{\watt}}. \textbf{(d)} A Hahn-echo sequence is used to probe the spin coherence $T_{2,\text{s}}$. The decay of the echo signal yields $T_{2,\text{s}}=\SI{613}{\micro\second}$. \textbf{(e)} The spin coherence time is extended to \SI{2}{\milli\second} by CPMG dynamical decoupling using 8 decoupling pulses. \textbf{(f)} The CPMG measurement was conducted for $N=1, 2, 4$ and 8 refocusing pulses, showing an increase in coherence time by a factor $N^{\beta}$ with $\beta=$ \SI{0,53\pm 0,03}.} 
	\label{fig:figure4} 
\end{figure}

Finally, we harness the optical spin initialization and readout to investigate coherent nuclear spin control. As a first demonstration, we perform nuclear Rabi oscillations, and adapt the pulse sequence for ODNMR by choosing a resonant RF frequency to match the \SI{21,5}{\mega\hertz} transition and vary the pulse length in steps of \SI{1}{\micro\second}. Fig.~\ref{fig:figure4}a shows the resulting Rabi oscillations, which reveal a Rabi frequency $\Omega_R = \SI{14}{\kilo\hertz}$ at an RF power of \SI{92}{\watt}. The damping of the oscillation originates from the inhomogeneity of the transition. We repeat this measurement for different RF powers and observe oscillations with increasing Rabi frequency, see Fig.~\ref{fig:figure4}b. We find a power dependence that follows the expected square root law $\Omega_R\propto \sqrt{P}$ with a proportionality factor of \SI{1,48}{\kilo\hertz\per\sqrt{\watt}}, see Fig.~\ref{fig:figure4}c.

The inhomogeneous broadening of the spin transition as well as slow fluctuations of the local magnetic field lead to dephasing that can be compensated by pulsed NMR sequences. We implement a Hahn-echo sequence, using a $\pi$ pulse duration of \SI{36}{\micro\second} for $\Omega_R=\SI{14}{\kilo\hertz}$ as obtained from Rabi oscillations. In order to obtain a quantitative contrast, the sequence is performed twice, once with and once without a phase shift of $\SI{180}{\degree}$ of the final $\pi/2$ pulse. The difference between the two measurements is normalized to their sum and referred to as the visibility. Fig.~\ref{fig:figure4}d shows an exemplary dataset of the visibility as a function of the delay time $\tau$, yielding an exponential decay with a time constant corresponding to the coherence time $T_{2,\text{s}}=\SI{0,61 \pm 0,04}{\milli\second}$. This value is comparable to the spin coherence observed in high-quality europium-doped solid-state crystals \cite{goldner_chapter_2015}, and underlines the promising properties of this molecular material. 

To further protect the nuclear spins from decoherence, dynamical decoupling control using Carr-Purcell-Meiboom-Gill (CPMG) sequences was applied. Such protection can, under certain circumstances, significantly increase the spin coherence time, as this sequence also compensates for pulse imperfections. This is achieved by applying the refocusing $\pi$ pulses along a $\ang{90}$-rotated axis compared to the $\pi/2$ pulses. CPMG measurements were performed for $N=1, 2, 4$ and 8 refocusing pulses. Fig.~\ref{fig:figure4}e depicts a representative measurement for $N=8$ pulses, for which we could observe a spin coherence time of up to \SI{2 \pm 0.2}{\milli\second}. The coherence time follows a scaling law described by

\begin{equation}
T_{2,\text{s(CPMG)}}=T_{2,\text{s(Echo)}}\cdot N^{\beta}\quad,
\end{equation}

where $\beta$ denotes the scaling factor and $T_{2,\text{s(Echo)}}$ the coherence time obtained from Hahn-echo measurements. All visibility decays were fitted with stretched exponential functions (stretching factors 1.43, 1.18, 1.32 and 1.31 for 1, 2, 4, and 8 $\pi$ pulses, respectively). The results, illustrated in Fig.~\ref{fig:figure4}f, demonstrate that dynamical decoupling significantly improves the spin coherence time, with a scaling factor of \SI{0,53 \pm 0,03}{}, which is slightly off the $2/3$-scaling expected for a correlated noise bath of a single spin species \cite{Dobrovitski_2012,Lukin_Walsworth_2012}. The scaling shows no saturation, such that increasing the number of refocusing pulses could extend the coherence time further. 
The coherence extension indicates that one dominating noise source originates from a bath with a correlation time that is long compared to the echo coherence time. When analyzing the obtained coherence values in the framework of an Ornstein-Uhlenbeck model for the bath \cite{wang_comparison_2012,Pascual-Winter2012}, we estimate a bath correlation time $\tau_B \approx \SI{13}{\milli\second}$ (see supplementary information) and a bath coupling strength $b \approx \SI{12}{\kilo\hertz}$ obtained from fitting the most narrow spin hole observed with a Gaussian function. One expected origin of the bath noise is fluctuating proton nuclear spins at the ligand. However, the coupling strength $b$ is larger than expected from nearby nuclear spins (few hundred Hertz), pointing towards paramagnetic impurities as a possible additional contribution. Indeed, we expect the presence of other lanthanide species according to the purity level of \SI{99.99}{\percent} of the europium salt precursor, such as Gd$^{3+}$, Nd$^{3+}$, and Dy$^{3+}$, that have comparably large magnetic moments. Furthermore, also quasi-localized low-frequency vibrational modes that couple to the hyperfine states may contribute to spin dephasing at the temperature used here \cite{kozankiewicz_single-molecule_2014,serrano_ultra_2022}. 

\section{Conclusion}\label{sec:conclusion}
In summary, our results have shown direct access to coherently controlled nuclear spins in a molecular material. Due to the weak magnetic moment of the europium nuclear spin, a long spin coherence could be observed, and even longer coherence is expected by applying more decoupling pulses and operating at millikelvin temperatures in a magnetic field to polarize paramagnetic impurities and freeze low-frequency vibrational modes. Also, purification and chemical engineering of the complex, e.g. by deuteration, is expected to improve nuclear spin coherence. Furthermore, optical addressing of nuclear spins may open a new avenue into NMR-based materials characterization. Super-hyperfine coupling to neighboring spins may offer unique signatures that could allow for structure analysis \cite{karlsson_nuclear_2017,abobeih_atomic-scale_2019}, enable polarization of ligand-centered nuclear spins, and possibly their full quantum control. Further, the coherent optical transitions are also a powerful tool to achieve direct optical spin manipulation \cite{rippe_experimental_2008,serrano_all-optical_2018}. This holds great promise for realizing fast spin qubit control including single- and two-qubit gates \cite{kinos_designing_2021}. When studied at the single-molecule level, e.g. by integration into nanophotonic cavities \cite{chen_parallel_2020,ruskuc_multiplexed_2025,toninelli_single_2021}, optically addressable nuclear spins in molecules offer a promising route to realize atomically precise multi-qubit quantum registers for scalable and optically connectable quantum processing nodes \cite{kinos_high-connectivity_2022,ruskuc_multiplexed_2025}.

\newpage

\section{Figures}\label{sec:figures}

\clearpage


\section*{Acknowledgments}
We acknowledge helpful discussions with Jannis Hessenauer, Ioannis Karapatzakis and Robin Wittmann.
\paragraph*{Funding:}
This work was funded by the Deutsche Forschungsgemeinschaft (DFG) through the Collaborative Research Centre “4f for Future” (CRC 1573 project number
471424360, project C2), the Max-Planck-School of Photonics, and the Karlsruhe School of Optics and Photonics (KSOP).
\paragraph*{Author contributions:}
E.V. and V. U. Ch. set up the experiment and conducted the measurements. S. K. synthesized the complexes. J. R. and D. S. guided experimental protocols. P. G., M. R. and D. H. advised on all efforts. All authors contributed to the data analysis and the manuscript preparation.
\paragraph*{Competing interests:}
There are no competing interests to declare.
\paragraph*{Data and materials availability:}
The data is published on the repository zenodo.org.


\section*{Supplementary materials}
Materials and Methods\\
Theoretical considerations\\
Figs. S1 to S3\\
Tables S1\\

\newpage


\clearpage

\bibliography{EuNuclearSpinControl}

\end{document}


\title[Supplementary Material for
Optically detected nuclear magnetic resonance of coherent spins
in a molecular complex]{Supplementary Material for
Optically detected nuclear magnetic resonance of coherent spins
in a molecular complex}


\author*[1,2]{\fnm{Evgenij} \sur{Vasilenko}}\email{evgenij.vasilenko@kit.edu}
\equalcont{These authors contributed equally to this work.}

\author[2]{\fnm{Vishnu} \sur{Unni Chorakkunnath}}\email{vishnu.chorakkunnath@kit.edu}
\equalcont{These authors contributed equally to this work.}

\author[2]{\fnm{Jeremias} \sur{Resch}}

\author[2]{\fnm{Nicholas} \sur{Jobbitt}}

\author[3]{\fnm{Diana} \sur{Serrano}}

\author[3]{\fnm{Philippe} \sur{Goldner}}

\author[1]{\fnm{Senthil Kumar} \sur{Kuppusamy}}

\author[1,4,5]{\fnm{Mario} \sur{Ruben}}

\author*[1,2]{\fnm{David} \sur{Hunger}}\email{david.hunger@kit.edu}

\affil*[1]{\orgdiv{Institute for Quantum Materials and technologies (IQMT)}, \orgname{Karlsruhe Institute of Technology}, \orgaddress{\city{Karlsruhe}, \postcode{76131},\country{Germany}}}

\affil*[2]{\orgdiv{Physics Institute (PHI)}, \orgname{Karlsruhe Institute of Technology}, \orgaddress{\city{Karlsruhe}, \postcode{76131},\country{Germany}}}

\affil[3]{\orgdiv{Chimie ParisTech}, \orgname{PSL University, CNRS, Institut de Recherche de Chimie Paris}, \orgaddress{\city{Paris}, \postcode{75231}, \country{France}}}

\affil[4]{\orgdiv{Institute of Nanotechnology (INT)}, \orgname{Karlsruhe Institute of Technology}, \orgaddress{\city{Karlsruhe}, \postcode{76131}, \country{Germany}}}

\affil[5]{\orgdiv{Centre Européen de Sciences Quantiques (CESQ)}, \orgname{Institut de Science et d'Ingénierie Supramoléculaires (ISIS)}, \orgaddress{\city{Strasbourg}, \postcode{67083}, \country{France}}}

\maketitle

\renewcommand{\thefigure}{S\arabic{figure}}
\renewcommand{\thetable}{S\arabic{table}}
\renewcommand{\theequation}{S\arabic{equation}}
\renewcommand{\thepage}{S\arabic{page}}
\setcounter{figure}{0}
\setcounter{table}{0}
\setcounter{equation}{0}
\setcounter{page}{1}

\subsubsection*{This PDF file includes:}
Materials and Methods\\
Theoretical considerations\\
Figures S1 to S3\\
Table S1\\

\newpage


\subsection*{Materials and Methods}

\subsubsection*{Growth of molecular crystals}
\label{subsubsec:mc_growth}
The all-oxygen coordination environment of the $\mathrm{Eu}^{3+}$ optical center is described as biaugmented trigonal prism and exhibits a $C_{2v}$ point group symmetry. We grow macroscopic molecular crystals by using the method of slow solvent evaporation in order to produce millimeter-sized large molecular crystals. In this method, the crystalline powder sample is dissolved in a small amount (\SI{100}{\milli\gram} in \SI{6}{\milli\litre}) of ethanol to reach a near-saturated solution. The solvent slowly evaporates over time, while macroscopic crystals start to form and slowly grow. Depending on the solvent evaporation rate, the quality of the crystal varies drastically. A rough control of the evaporation rate is achieved by perforating the covering foil of the beaker with the solution. We present results on two different macroscopic crystals. The first crystal was grown with a solvent evaporation time close to three weeks. This crystal has lateral dimensions of about $\SI{1}{\milli\m}\times\SI{1}{\milli\m}$ and a thickness of approximately \SI{200}{\micro\m}. The crystal was integrated into a dilution refrigerator. In order to provide efficient thermalization, the crystal was glued onto a copper sheet and positioned on a copper mount for free space excitation with a confocal microscope. Only the inhomogeneous linewidth was measured for this crystal under these conditions. The second crystal was recrystallized within a time period of four days with dimensions of $\SI{1,5}{\milli\m}\times\SI{1,5}{\milli\m}$ and a thickness of about \SI{500}{\micro\m}. This crystal was integrated into a fiber-based ferrule setup, and directly immersed in liquid helium, providing effective thermalization and a stable temperature of \SI{4.2}{\kelvin}. All other measurements shown were preformed with the second crystal under these conditions.

A superconducting coil, which is described in detail in Subsection \textbf{RF setup}, was installed around the ferrule setup and combined with an LC-circuit to enable the addressing of both hyperfine transitions of the molecular crystals (see Fig.~\ref{subsubsec:rf_coil}a). Both crystals were grown from dissolved microcrystalline powder with natural abundance of the Europium isotopes with \SI{100}{\percent} $\mathrm{Eu}^{3+}$ ion doping concentration. All measurements were performed on the $^{151}\mathrm{Eu}^{3+}$ isotope, since the reported hyperfine splittings shown in Fig.~1b are smaller compared to the other isotope \cite{serrano_ultra_2022}, and are therefore accessible with the frequency range provided by our setup.

\subsubsection*{Optical setup}
\label{subsubsec:opt_set}

For both cryostats used, the same optical setup could be employed. The molecular samples were resonantly excited by a Sirah Matisse 2DX dye laser, which exhibits a FWHM linewidth of $<$ \SI{50}{\kilo\hertz} and can be mode-hop-free tuned over (\SI{\sim 75}{\giga\hertz}). The wavelength for all experiments was adjusted to approximately \SI{580.377}{\nano\meter} to match the coherent $\mathrm{}^7\mathrm{F}_0\rightarrow\mathrm{}^5\mathrm{D}_0$ transition. The main laser beam is divided into two parts: one beam is used as a reference path, as the laser output is monitored via an optical spectrum analyzer (OSA, Bristol Instruments 771 Series) to verify the laser frequency and single-mode operation, while the other beam is guided through an acousto-optic modulator (AOM, Gooch\&Housego 3200-121) in the double-pass configuration enabling the shaping and control of optical pulses in amplitude and frequency. The AOM is controlled by an arbitrary waveform generator (AWG, Quantum Machnies OPX+). After the AOM, the light is fiber-coupled and can be directed either to the dilution refrigerator or a home-built dipstick cryostat which incorporates the fiber-based ferrule setup. The current ferrule setup consists of two opposing small cylindrical ferrules with a diameter of \SI{2,5}{\milli\meter}, one made out of ceramic and the other out of Teflon. The ceramic ferrule has a bore size of \SI{300}{\micro\meter} which is employed to integrate the excitation multimode fiber with a core diameter of \SI{200}{\micro\meter}. Since no ceramic ferrules with bore sizes of \SI{1,5}{\milli\meter} are commercially available, we machined a Teflon ferrule to incorporate a multimode fiber with a core diameter of \SI{1000}{\micro\meter}. This large core fiber was pulled back by about \SI{0,7}{\milli\meter}, introducing enough sample space for large-scaled crystalline samples. Both ferrules were put in a commercial ceramic mating sleeve connector and glued to provide more stability of the setup. In the dilution refrigerator, the macroscopic crystal was placed onto a copper mount and excited in a free-space geometry. The signal transmitted through the crystal was collected through a window of the cryostat. In both cases, the resulting signal was detected by a Thorlabs avalanche photodetector (APD, APD130A2/M). To enable selective detection of optical signals from the sample, additional filters can be employed. A longpass filter (BLP01-594R-25) is used to block the excitation light and observe fluorescence, while a bandpass filter (FB580-10) with a spectral range of \SI{\pm 10}{\nano\meter} is used for detecting absorption around the resonantly excited $\mathrm{}^7\mathrm{F}_0\rightarrow\mathrm{}^5\mathrm{D}_0$ transition.  

\subsubsection*{RF setup}
\label{subsubsec:rf_coil}

The RF signal is produced by the same AWG used for the creation of the optical pulses, allowing precise control and synchronization between applied optical and RF pulses. The RF signal is guided through an \SI{10}{\decibel} attenuator to prevent damage from potential reflections, and then amplified using a high-power RF amplifier (Bonn Elektronik, BSA 1025-150) capable of delivering up to \SI{150}{\watt} of output power. The amplified signal is coupled into the cylindrical RF coil via a capacitor-based \SI{50}{\ohm} matching circuit, which allows tuning of the resonance frequency, bandwidth, and depth of the resonance through adjustable parallel and series capacitors (a few tens of \SI{}{\pico\farad}). Before a measurement, a vector network analyzer (VNA, E5071B ENA) is connected to this LC-circuit in order to monitor the resonance profile and to adjust the resonance frequency to the specific hyperfine transition. On the output side of the RF coil, a high-power \SI{30}{\decibel} attenuator is placed to reduce these high powers significantly, which is followed by additional attenuators and a \SI{50}{\ohm} termination connector to suppress reflections and ensure impedance matching.

For the measurements presented in this work, a home-built cylindrical coil was used. The coil features closely spaced turns without air gaps and is encapsulated in epoxy adhesive to ensure mechanical stability. The following paragraph summarizes its key characteristics in detail.

 \begin{itemize}
     \item \textbf{Wire material}: copper-stabilized superconducting wire from SUPERCON (type 54S43) and diameter of \SI{0,43}{\milli\meter} (insulated) 
     \item \textbf{Geometry:} see Fig.~\ref{fig:sup_coil}a and Table \ref{tab:coil_param} 
     \item \textbf{Characterization of resonance:} see Fig.~\ref{fig:sup_coil}b, FWHM of \SI{1.9}{\mega\hertz} and quality factor $Q \SI{\sim 11}{}$.
 \end{itemize}

\subsubsection*{Spectral pit preparation}
\label{subsubsec:pit_prep}

The spin characterization of the complex requires spin polarization of the $^{151}\mathrm{Eu}^{3+}$ nuclei. A general spin initialization by preparing a spectral pit is carried out in this work. The lifetime of the spectral pit is limited by spin $T_{1,\text{s}}$, which is orders of magnitude longer than the spin characterization measurement sequence time. The laser frequency is swept by \SI{10}{MHz} around the peak of the optical inhomogeneous lines within \SI{300}{ms}. This is repeated 20 times to prepare a spectral pit with a contrast of about \SI{60}{\percent}. An exemplary preparation sequence is shown in Fig.~\ref{fig:spectral_pit}a. The prepared spectral pit is shown in Fig.~\ref{fig:spectral_pit}b. This spectral pit contains nine classes of $^{151}\mathrm{Eu}^{3+}$ ions out of which six contribute to the spin signal when driven resonantly. After the pulse sequences for spin driving, all of which are finished within \SIrange{1}{15}{\milli\second}, a series of optical pulses with a frequency chirp over \SI{100}{MHz} within \SI{300}{ms} is applied 20 times to restore the spin states of $^{151}\mathrm{Eu}^{3+}$ ions to their initial distribution.

\newpage


\subsection*{Theoretical considerations}
\label{subsec:theo_consid}

\subsubsection*{Addressable ions in a spin measurement}
\label{subsec:number_ions}
In the following, we estimate the order of magnitude of ions that are resonantly driven by the RF field and optically readout in a single measurement, taking into account our experimental conditions such as sample geometry, pulse parameters, and measured signals.

We assume that the \SI{200}{\micro\meter} core multimode fiber ($\mathrm{NA}=\SI{0.5}{}$) is in contact with the flat surface of the macroscopic crystal, which is integrated into the Teflon ferrule. The propagation path is taken as the \SI{500}{\micro\meter} thickness of the crystal, and the refractive index of the crystalline material is \SI{1.5}{} \cite{serrano_ultra_2022}. The excitation volume through the crystal is approximated as a truncated cone with an initial diameter of \SI{200}{\micro\meter} and an exit exit diameter of \SI{\sim 400}{\micro\meter}, leading to an illuminated volume of $V$ \SI{\sim 10e-4}{\cubic\cm}. Given the concentration of $C_{\text{Eu}} = \SI{9.6e20}{\ionspercubecm}$, the number of ions spatially located within the light cone is \SI{\sim 1e17}{\ions}. 
We account for the fraction of ions optically probed within the inhomogeneous line by $\eta_\text{h}=\Gamma_\text{h}/\Gamma_\text{inh}\approx 1/23000$, the fraction of the $^{151}\mathrm{Eu}^{3+}$ isotope $\eta_{151}=0.5$, and the fraction of ions within the spin inhomogeneous line driven by the RF field including the overall spin contrast $\eta_{\text{s}}=0.2\times \Gamma_{\text{s,h}}/\Gamma_{\text{s,inh}}$, and obtain a probed ion number
\begin{equation}
    N_p= C_\text{Eu}V\eta_\text{h}\eta_{151}\eta_\text{s}\approx 10^{10}.
\end{equation}
With a collection efficiency of $\lesssim \SI{1}{\percent}$, we thus detect signals corresponding to $\SI{\sim e8}{}$ ions. Using a single photon counting module instead of an APD, about $\SI{e5}{}\times$ smaller signals can be recorded with similar signal to noise ratio. With improved photonic structures for higher collection efficiency, this will enable ODNMR on nanoscopic ensembles.

\subsubsection*{Nuclear spin dephasing sources}
\label{subsec:nuclear_nuclear_int}

One dephasing source for the europium nuclear spin is expected to be surrounding nuclear spins at the ligands. In order to extract a bath correlation time and a bath coupling strength, we assumed an Ornstein-Uhlenbeck model for the bath and fitted the decay of the visibility in CPMG measurements according to the following relation \cite{Pascual-Winter2012}: 
\begin{equation}\label{eq:formula_fit}
\overline{\rho}(N,\tau)=a\cdot\exp\left(-(\sigma \tau_c)^2
\left\{
\left[
\frac{1}{\tau_c} - \frac{2}{\tau} \tanh\left( \frac{\tau}{2\tau_c} \right)
\right] t
- \left[ 1 + (-1)^{N+1} e^{-t/\tau_c} \right]
\left[ 1 - \operatorname{sech}\left( \frac{\tau}{2\tau_c} \right) \right]^2
\right\}\right),
\end{equation}
where $t=N\cdot\tau$ represents the total evolution time, $N$ denotes the number of applied refocusing pulses during the CPMG measurements, $\sigma$ is the bath coupling strength and $\tau_c$ the bath correlation time. To assess whether the value of the bath coupling strength obtained from spin hole burning is consistent with a bath of nuclear spins located on the ligands, we calculate the interaction strength between a single europium nuclear spin and a hydrogen nuclear spin. The interaction strength is given by:
\begin{equation}\label{eq:coupling}
    E=\frac{\mu_0}{4\pi}\cdot\frac{\mu_{\text{Eu}}\mu_{\text{H}}}{r^3}.
\end{equation}

The molecular complex features 48 hydrogen atoms and a nearby nitrogen atom (\SI{\sim 4}{\angstrom}), which is located in the counter ion. The distance for the hydrogen atoms ranges between \SI{4}{\angstrom} to \SI{8}{\angstrom}. The gyromagnetic ratio of europium is \SI{6.65e7}{\radian\per\second\per\tesla} ($\mathrm{I}=5/2$) and for a proton \SI{2.68e8}{\radian\per\second\per\tesla} ($\mathrm{I}=1/2$), leading to an interaction strength of \SI{\approx 72}{\hertz} at \SI{8}{\angstrom} and \SI{\sim 583}{\hertz} at \SI{4}{\angstrom}. Both calculated coupling values are small compared to the extracted \SI{12}{\kilo\hertz} bath coupling strength, and also the total field of a randomly oriented nuclear spin ensemble is expected to lead to a smaller coupling. 

As an additional noise source, other REI species originating from impurities in the precursor materials can exhibit an electron spin magnetic moment that would couple stronger to the europium nuclear spin.
For an electron spin contribution, we assume the magnetic moment of $\mu_{\text{B}}$ ($g$-factor of 2). The formula \ref{eq:coupling} can be rearranged to solve for the distance $r$. By inserting the coupling strength of \SI{\approx 12}{\kilo\hertz} into the equation, we obtain a distance of \SI{\approx 13}{\angstrom}. This resulting distance is comparable to the separation of neighboring Eu centers within the unit cell, which range from \SIrange{9,35}{10,40}{\angstrom}. However, the impurity concentration of the europium chloride used for synthesizing the molecular complex was about \SI{99,99}{\percent}, and average impurity distances are expected to be \SI{\sim 20}{\nano\meter}. Impurities are thus not expected to dominate the dephasing.

Finally, quasi-local low-frequency vibrational modes \cite{kozankiewicz_single-molecule_2014} have been observed to affect the optical homogeneous linewidth in the studied complex above \SI{3.5}{\kelvin} \cite{serrano_ultra_2022}. Such vibrations can modulate the local electric field gradient at the Eu nucleus, which affects the ligand field contribution to the quadrupole splitting. This is expected from the observed correlation of the spin and optical transition frequency. Already a moderate reduction in temperature may significantly reduce this contribution.


\newpage

\subsection*{Figures}\label{sec:SI_figures}

\begin{figure}[h] 
	\centering
	\includegraphics[width=\textwidth]{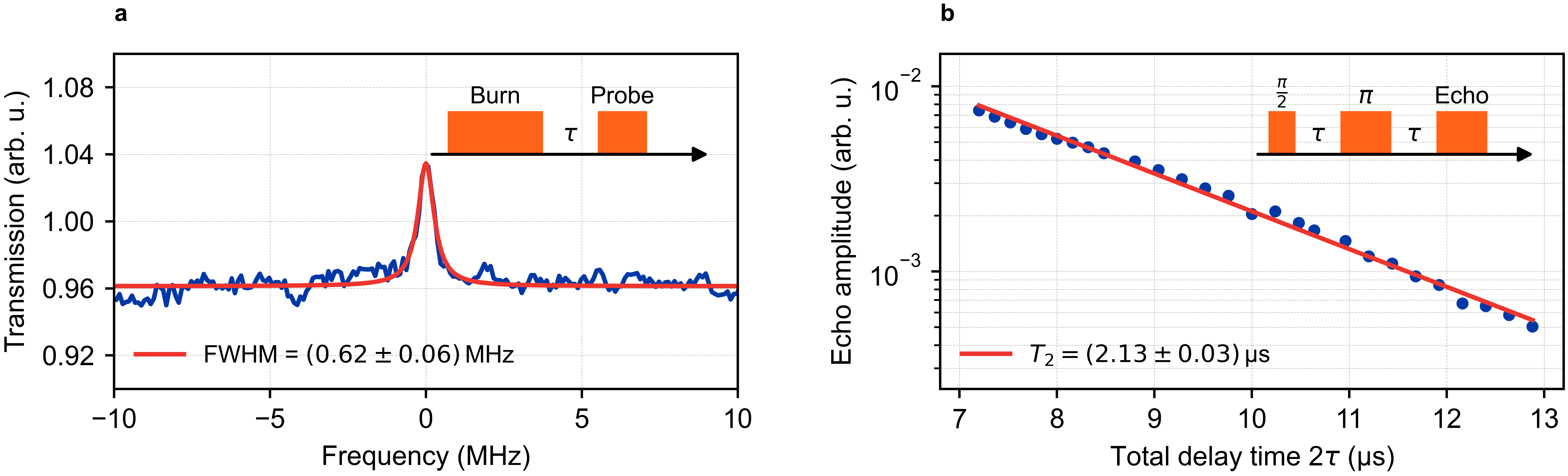} 

	\caption{\textbf{Characterization of optical properties. (a)} A spectral hole burnt into the inhomogeneous profile of crystal 2 yields a hole linewidth of \SI{620 \pm 60}{\kilo\hertz}. \textbf{(b)} Photon echo amplitude decay as a function of the total delay time $2\tau$, resulting in an optical coherence time $T_{2,\text{o}}$ of \SI{2.13 \pm 0.03}{\micro\second}. The echo signal was measured using heterodyne detection. Both measurements were performed at a temperature of \SI{4.2}{\kelvin}.}
	\label{fig:sup_optical_charac} 
\end{figure}

\begin{figure}[h]
	\centering
	\includegraphics[width=\textwidth]{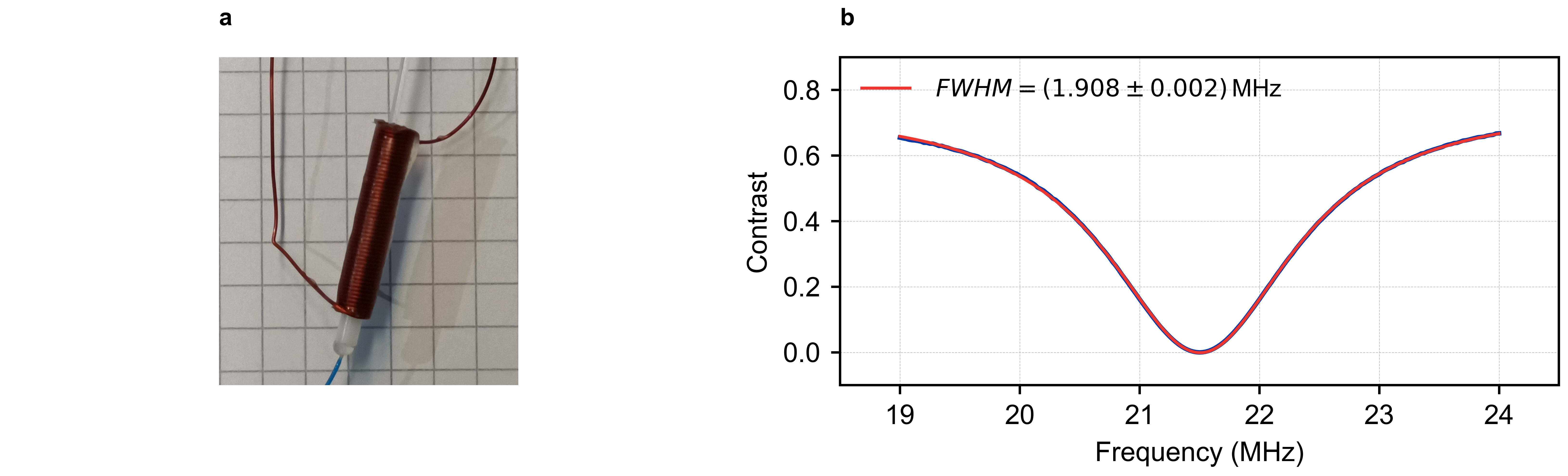} 
	\caption{\textbf{Resonance characterization of the RF coil. (a)} A photograph of the experimental setup featuring both excitation (blue jacket) and collection fiber (transparent) covered by the RF coil with compact winding geometry. The coil is encapsulated by an epoxy adhesive to ensure mechanical robustness. \textbf{(b)} Reflection spectrum of the same coil, tuned to \SI{21.5}{\mega\hertz}. A Fano resonance fit is applied to extract the FWHM of \SI{\sim 1.9}{\mega\hertz}, yielding a quality factor $Q \SI{\sim 11}{}$.}
	\label{fig:sup_coil} 
\end{figure}

\begin{figure}[h]
	\centering
	\includegraphics[width=\textwidth]{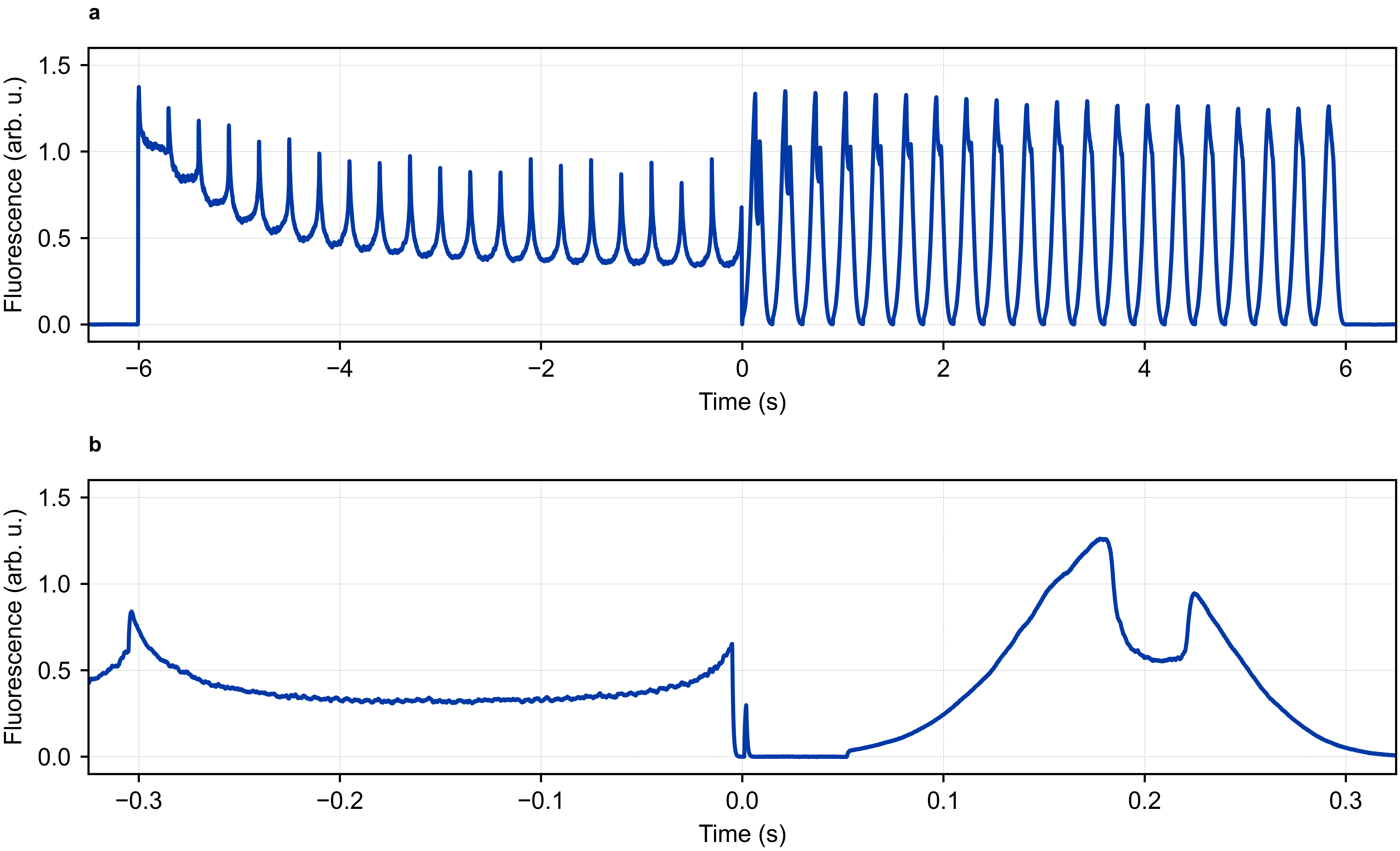} 

	\caption{\textbf{Spectral pit preparation. (a)} Full time trace of one exemplary spectral pit preparation sequence. \textbf{(b)} Zoom into the spectral pit preparation sequence showing the final pit burn, the short probe pulse and the start of the erasing sequence.}
	\label{fig:spectral_pit} 
\end{figure}

\clearpage

\subsection*{Tables}\label{sec:SI_tables}

\begin{table}[h] 
	\centering
	\caption{\textbf{Coil geometry.}
		}
	\label{tab:coil_param} 

	\begin{tabular}{lc} 
		\\
		\hline
		Parameter & Used coil\\
		\hline
		Number of turns & 62\\
		Inner diameter (mm) & 2.55 \\
		Outer diameter (mm) & 3.45 \\
        Length (cm) & 2\\
        
		\hline
	\end{tabular}
\end{table}

\newpage


\clearpage

\bibliography{EuNuclearSpin_SI}